\documentclass[aps,twocolumn,showkeys,amsmath,amssymb,superscriptaddress,floatfix]{revtex4}
\usepackage{upgreek}
\usepackage{graphicx}
\usepackage{color}
\newcommand{\rtrim}{$R\overline{3}m$}
\newcommand{\st}{$\rm ^\circ$}     
\newcommand{\stc}{$\rm ^\circ C$}  
\newcommand{\stt}{$\rm ^\circ 2\theta$}  

\newcommand{\muvk}{$\mu$V/K}

\newcommand{\emug}{emu/g}  
\newcommand{\mub}{$\mu_B$}

\newcommand{\bamhexfe}{BaFe$_{12}$O$_{19}$}

\newcommand{\yhexzn}{Ba$_2$Zn$_{2}$Fe$_{12}$O$_{22}$}
\newcommand{\yhexco}{Ba$_2$Co$_{2}$Fe$_{12}$O$_{22}$}
\newcommand{\basryhexzn}{Ba$_{2-x}$Sr$_x$Zn$_{2}$Fe$_{12}$O$_{22}$}

\newcommand{\yig}{Y$_{3}$Fe$_{5}$O$_{12}$}

\begin{document}
\sloppy
\title{Spin Seebeck effect in Y-type hexagonal ferrite thin films}
\author{J. Hirschner}
\author{K. Kn\'{\i}\v{z}ek}
\email[corresponding author: ]{knizek@fzu.cz}
\author{M. Mary\v{s}ko}
\author{J. Hejtm\'{a}nek}
\affiliation{
 Institute of Physics ASCR, Cukrovarnick\'{a} 10, 162 00 Prague 6, Czech Republic.}
\author{R. Uhreck\'{y}}
\author{M. Soroka}
\author{J. Bur\v{s}\'{\i}k}
\affiliation{
 Institute of Inorganic Chemistry ASCR, 250 68 \v{R}e\v{z} near Prague, Czech Republic.}
\author{A. Anad\'{o}n Barcelona}
\author{M. H. Aguirre}
\affiliation{
 Instituto de Nanociencia de Arag\'{o}n, Universidad de Zaragoza, E-50018 Zaragoza, Spain}
\begin{abstract}
Spin Seebeck effect (SSE) has been investigated in thin films of two Y-hexagonal ferrites Ba$_2$Zn$_{2}$Fe$_{12}$O$_{22}$ (Zn2Y) and Ba$_2$Co$_{2}$Fe$_{12}$O$_{22}$ (Co2Y) deposited by a spin-coating method on SrTiO$_3$(111) substrate. The selected hexagonal ferrites are both ferrimagnetic with similar magnetic moments at room temperature and both exhibit easy magnetization plane normal to $c$-axis. Despite that, SSE signal was only observed for Zn2Y, whereas no significant SSE signal was detected for Co2Y. We tentatively explain this different behavior by a presence of two different magnetic ions in Co2Y, whose random distribution over octahedral sites interferes the long range ordering and enhances the Gilbert damping constant. The temperature dependence of SSE for Zn2Y was measured and analyzed with regard to the heat flux and temperature gradient relevant to the SSE signal. 
\end{abstract}
%
%
\maketitle
\section{Introduction}


Spintronics is a multidisciplinary field which involves the study of active manipulation of spin degrees of freedom in solid-state systems \cite{xu2015handbook}. Thermoelectricity concerns the ability of a given material to produce voltage when temperature gradient is present, thus converting thermal energy to electric energy \cite{rowe2005thermoelectrics}. The emerging research field of spin caloritronics, which may be regarded as interconnection of spintronics and thermoelectricity, combines spin-dependent charge transport with energy or heat transport.
One of the core elements of spin caloritronics is the spin-Seebeck effect discovered in 2008 by Uchida \textit{et al.} \cite{RefUchida2008NAT455_778}. The spin-Seebeck effect (SSE) is a combination of two phenomena - the generation of a spin current by a temperature gradient applied across a magnetic material, and a conversion of the spin current to electrical current by means of the inverse spin Hall effect (ISHE) \cite{RefSaitoh2006APL88_182509} in the attached metallic thin layer. A necessary condition for the observation of SSE is that the directions of the spin current, magnetic moments of the magnetic material, and electrical current in the metallic layer, are mutually perpendicular.
Since the resulting electric field is related to temperature gradient, it is possible in the regime of linear response to define a spin Seebeck coefficient $S_{SSE} = E_{ISHE} / \nabla T$.


As regards the magnetic material as a source of the spin current, it is more convenient to use insulators rather than conductors, in order to avoid parasitic signals such as a planar or anomalous Nernst effect \cite{RefUchida2010NATMAT9_894}. There are three main types of magnetic insulators possessing critical temperature $T_C$ above the room temperature: garnets, spinels, and hexagonal ferrites. So far, most of the SSE experiments employed iron-based garnet because of their very low Gilbert damping constant, \textit{i.e.} slow decay of spin waves, since this decay limits the thickness of the magnetic layer that actively generates the spin flow.


In this work we have focused on Y-hexaferrites as magnetic material, namely \yhexzn\ (Zn2Y) and \yhexco\ (Co2Y). Their mass magnetizations at room temperature are 42.0~\emug\ for Zn2Y and 34~\emug\ for Co2Y \cite{RefPullar2012PROGMS57_1191}, which are higher than 27.6~\emug\ of yttrium ferrite garnet \yig\ \cite{RefNIST2853}. Since a positive correlation between SSE and the saturation magnetization has also been proposed \cite{RefUchida2013PRB87_104412}, Y-hexaferrites appear to be a suitable material for the spin current generation in the spin-Seebeck effect.

\begin{figure}
\centering
\includegraphics[width=0.90\columnwidth,viewport=45 410 535 780,clip]{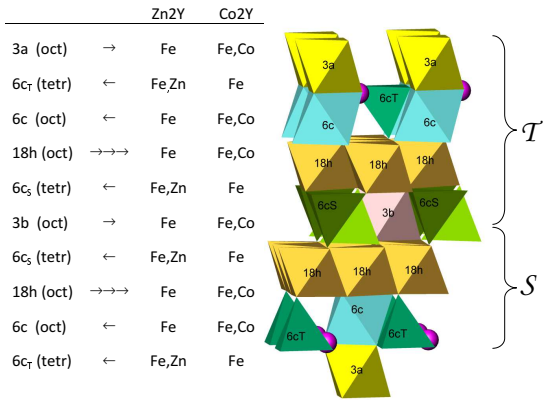} 
\caption{One formula unit of \yhexzn\ or \yhexco\ structure with alternating structural blocks $S$ and $T$. Shown are Fe, Co and Zn polyhedra and Ba cations (magenta bullets). The description includes Wyckoff positions, types of polyhedra (tetrahedral or octahedral), arrows indicating spin direction of the collinear ferrimagnetic structure, and the preferential occupation of sites.} \label{Fig_Yhex}
\end{figure}

The crystal structure of Y-hexaferrites belongs to the trigonal space group \rtrim\ and is composed of alternating stacks of S (spinel Me$_2$Fe$_4$O$_8$, Me~=~Zn or Co in our case) and T (Ba$_2$Fe$_8$O$_{14}$) blocks along the hexagonal $c$-axis. The magnetic configuration of Y-hexaferrites is usually ferrimagnetic, with spin up orientation in octahedral $3a$, $3b$ and $18h$ sites and spin down in tetrahedral $6c_T$, $6c_S$ and octahedral $6c$ sites, see Fig.~\ref{Fig_Yhex}.

Magnetocrystalline anisotropy is observed in all hexagonal ferrites, which means that their induced magnetisation has a preferred orientation within the crystal structure,
either with an easy axis of magnetisation in the $c$-direction or with an easy plane of magnetisation perpendicular to $c$-direction, the latter being the case of the selected Y-hexaferrites.
Due to their direction of easy grow lying in $ab$-plane, hexaferrites inherently tend to grow with their $c$-axis perpendicular to the film plane when deposited as thin films. Since the magnetization vector in SSE element should lie in parallel to the film surface, the hexaferrites with an easy plane of magnetisation are more suitable for the SSE experiment.

The principal difference between Zn2Y and Co2Y comes from a different site preferences and magnetic properties of Zn$^{2+}$ and Co$^{2+}$. Zn$^{2+}$ ion is non-magnetic ($d^{10}$) and occupies preferentially the tetrahedral sites. Since both Fe$^{3+}$ in tetrahedral sites have spin down orientation, the substitution of Zn$^{2+}$ to these sites maximize the overall magnetic moment and the saturation magnetisation at low temperature reaches 18.4~\mub\ (theoretical limit considering 5~\mub\ \textit{per} Fe$^{3+}$ would be 20~\mub). However, because of the relatively low critical temperature $T_C \sim 130$\stc, the magnetization at room temperature is only about 10.7~\mub\ (42~\emug). Co$^{2+}$ ion is in the low spin state (LS, $t_{2g}^6e_g^1$) and occupies preferentially the octahedral sites. The resulting magnetic moment depends on the actual distribution of Co between octahedral sites occupied by Fe$^{3+}$ with spin up or spin down orientation, nevertheless generally will be much lower than in Zn2Y and the typical saturation magnetization is around 10~\mub. On the other hand, since the critical temperature $T_C \sim 340$\stc\ of Co2Y is higher, the magnetic moment at $T_{room}$ around 8.6~\mub\ is not so different from that of Zn2Y, see \textit{e.g.} the review paper \cite{RefPullar2012PROGMS57_1191}.

Spin-Seebeck effect in Y-hexaferrite was studied for the compound of stoichiometry \basryhexzn\ ($x=1.5$) \cite{RefTakagi2016APLMAT4_32502}. In this study it was observed, that the magnitude of SEE is proportional to bulk magnetization even through the successive magnetic transitions among various helimagnetic and ferrimagnetic phases.
M-type hexaferrite \bamhexfe\ was studied in \cite{RefLi2014APL105_242412}. Since M-type hexaferrite have strong anisotropy with an easy axis of magnetisation in the $c$-direction, a proper substrate and deposition procedure must be selected in order to grow the thin films with the $c$-axis oriented parallel to the surface. The advantage of M-type is its high coercive field, which makes the resulting SSE element self-biased, thus producing SEE signal even without presence of magnetic field.
Spin-Seebeck effect was also studied in Fe$_3$O$_4$ with spinel structure, which may be in some context considered as the simplest structural type of hexagonal ferrites. Large coercive fields and high saturation magnetisation makes Fe$_3$O$_4$ promising magnetic material for the investigation of self-biased SSE elements
\cite{RefRamos2013APL102_72413,RefWu2014APL105_92409,RefRamos2015PRB92_220407,RefAnadon2016APL109_12404,RefCaruana2016PPSRRL10_613,RefRamos2016APLMAT4_104802}


Current research describes the SSE using typical quantity of spin Seebeck coefficient with unit of \muvk, which is in conventional thermoelectric materials used for evaluating the effectiveness of the process. However, in most of the experimental setups the temperature sensors measuring the temperature difference $\triangle T$ are attached to the measure cell itself. This implies that $\triangle T$ describes not only the thermal characteristics of the studied material, but the whole measurement cell instead, making the quantity in unit of $\mu V/K$ physically irrelevant to the spin Seebeck effect itself. This issue was studied in details in Ref.~\cite{RefSola2015JAP117_17C510}. The authors pointed out, that when using the setup dependent $\triangle T$ as independent variable the determined SSE can be hardly comparable between laboratories. In order to solve this problem, the authors designed a measurement system with precise measurement of the heat flux through the sample and proposed using heat flux or thermal gradient at the sample as the independent variable.

In this work we have followed this approach and manifested, that the total temperature difference $\triangle T$ is not suitable independent variable even for measuring within one setup if the temperature dependent experiment is performed, since the temperature evolution of thermal conductivity of the whole setup may be different from that of the sample material itself.

\section{Experimental}

\begin{figure}
\centering
\includegraphics[width=0.85\columnwidth,viewport=10 300 550 750,clip]{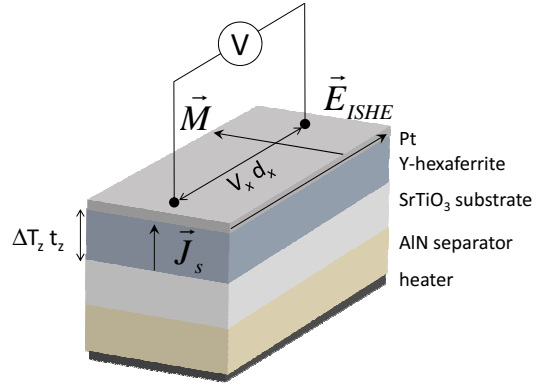} 
\caption{Schema of the longitudinal experimental configurations.
 Directions of temperature gradient ($\nabla T$), magnetization (M), spin current ($J_s$) and electrical field resulted from inverse spin Hall effect ($E_{ISHE}$) are shown. The meaning of parameters $V_{x}$, $t_{z}$, $d_{x}$ and $\triangle T_{z}$ used in eq.~\ref{eqssenorm} is also indicated.
} \label{Fig_ExpSSE}
\end{figure}


Thin films of \yhexzn\ (Zn2Y) and \yhexco\ (Co2Y) were prepared by spin-coating technique on (111)-oriented, epitaxially polished SrTiO$_3$ (STO) single crystals with metalorganic precursor solutions. Commercial 2-ethylhexanoates Me(CH$_3$(CH$_2$)$_3$CH(C$_2$H$_5$)COO)$_n$ ($n=2$ for Me~=~Ba, Co, Zn; n~=~3 for Me = Fe, ABCR, Germany) were used as precursors. Calculated amounts of metal precursors were dissolved in iso-butanol, mixed and heated for several hours at 80\stc\ to accomplish homogenization. Subsequently a suitable amount of 2,2-diethanolamine (DEA) used as a modifier was added. The modifier to alkali earth metal molar ratio was n(DEA)/n(alkali earth metal)~=~2. Prior to the deposition the stock solutions were usually diluted with iso-butanol to obtain films of desired thickness. All reactions and handling were done under dry nitrogen atmosphere to prevent reaction with air humidity and preliminary formation of alkaline earth carbonates in solutions. Single crystals of STO were washed in acetone combined with sonication and then annealed at 1200\stc\ in air for 24 hours to heal up the surface damage caused during polish treatment. Prior the deposition they were treated with plasma (Zepto Plasma cleaner, Diener Electronic, Germany). After the drying at 110\stc\ for several minutes and pyrolysis of gel films at 300\stc\ for 5~minutes, crystallization annealing was done at 1000\stc\ for 5~minutes in conventional tube furnace under open air atmosphere. The deposition-annealing cycle was repeated ten times to obtain the desired film with approximately $300-350$ nm of thickness.
Final annealing was done in tube furnace under open air atmosphere at 1050\stc\ for 5~min (Zn2Y) or 1000\stc\ for 60~min (Co2Y).



Spin Seebeck effect was measured using home-made apparatus. A longitudinal configuration was used, in which the directions of the spin current, magnetic moments and electrical current are mutually perpendicular \cite{RefUchida2014JPCM26_343202}, see Fig.~\ref{Fig_ExpSSE}. AlN plate with high thermal conductivity was used to separate the heater and the sample in order to uniformly spread the heat flux over the sample area. The thermal barriers between individual parts of the call were treated by appropriate greases (Apiezon type N, Dow Corning Varnish, Ted Pella silver paste).


The width of the measured sample was 2~mm, the length was 7~mm and contact distance was approx. 5~mm. Thickness of the Zn2Y-hexaferrite layers was between $300-350$~nm, the thickness for Co2Y-hexaferrite layers ranged between $150-300$~nm. Pt layer was deposited using K550X Quorum Technologies sputter coater. The thickness of the layer was determined by internal FTM detector (Tool factor 4.7), the final Pt deposition thickness was $\sim$8~nm. The resistance of the Pt-layer measured by a 2-point technique was within the range $350-650$~$\Omega$ at room temperature and linearly decreased by $10-15$\% down to 5~K, whereas the resistance of the Y-hexaferrite thin layer itself was more than G$\Omega$.
Therefore, the contribution from the anomalous Nernst effect (ANE) can be considered as negligible due to the resistivity difference between Y-hexaferrite and Pt layers.

The magnetic hysteresis loops were measured within the range of magnetic field from $-25$ to 25~kOe at room temperature using a SQUID magnetometer (MPMSXL, Quantum Design)

The phase purity and degree of preferred orientation of the thin films was checked by X-ray diffraction over the angular range $10-100$\stt\ using the X-ray powder diffractometer Bruker D8 Advance (CuK$\alpha_{1,2}$ radiation, secondary graphite monochromator).
Atomic force microscopy AFM (Explorer, Thermomicroscopes, USA) was used to evaluate surface microstructure of the thin films.


\section{Results and Discussion}


\begin{figure}
\centering
\includegraphics[width=0.95\columnwidth,viewport=0 350 600 800,clip]{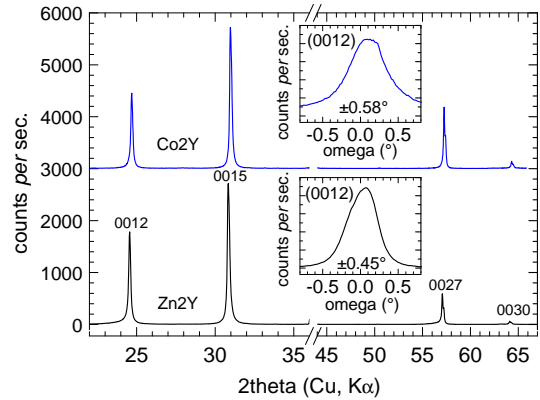} 
\caption{X-ray diffraction of the \yhexzn\ (black line, Zn2Y) and \yhexco\ (blue line, Co2Y) thin film. The insets show rocking-curve measurements. The diffraction peak (111) of the SrTiO$_3$ substrate is skipped.} \label{Fig_XRD}
\end{figure}

\begin{figure}
\centering
\includegraphics[width=0.85\columnwidth,viewport=100 520 500 780,clip]{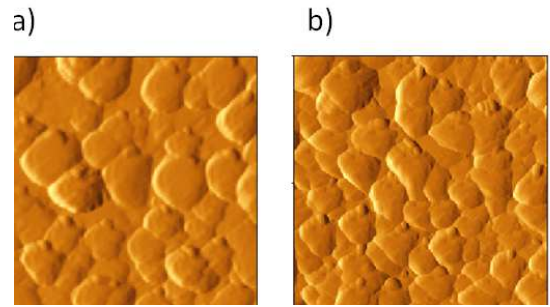} 
\caption{AFM images of surface topography of (a) \yhexzn\ (calculated roughness r.m.s.~=~27~nm) and (b) \yhexco\ (r.m.s~=~30~nm).} \label{Fig_AFM}
\end{figure}

The X-ray diffraction confirmed single phase purity of the thin film and $c$-axis preferred orientation, quantified by the full-width at the half-maximum (FWHM) of the rocking curve as 0.45\st\ for Zn2Y and 0.58\st\ for Co2Y, see Fig.~\ref{Fig_XRD}. The $c$-lattice parameters 43.567(7)~\AA\ for Zn2Y and 43.500(9)~\AA\ for Co2Y, calculated using $\cos\theta/\tan\theta$ extrapolation to correct a possible off-centre position of the film during XRD measurement, are in good agreement with literature values \cite{RefShin1993POWDIF8_98}.

Fig.~\ref{Fig_AFM} shows AFM images of surface topography of Zn2Y and Co2Y. Platelets with hexagonal shape can be identified in both images with similar shape and size. Calculated roughness (r.m.s.) values are around $27-30$~nm.


\begin{figure}
\centering
\includegraphics[width=0.85\columnwidth,viewport=0 0 772 591,clip]{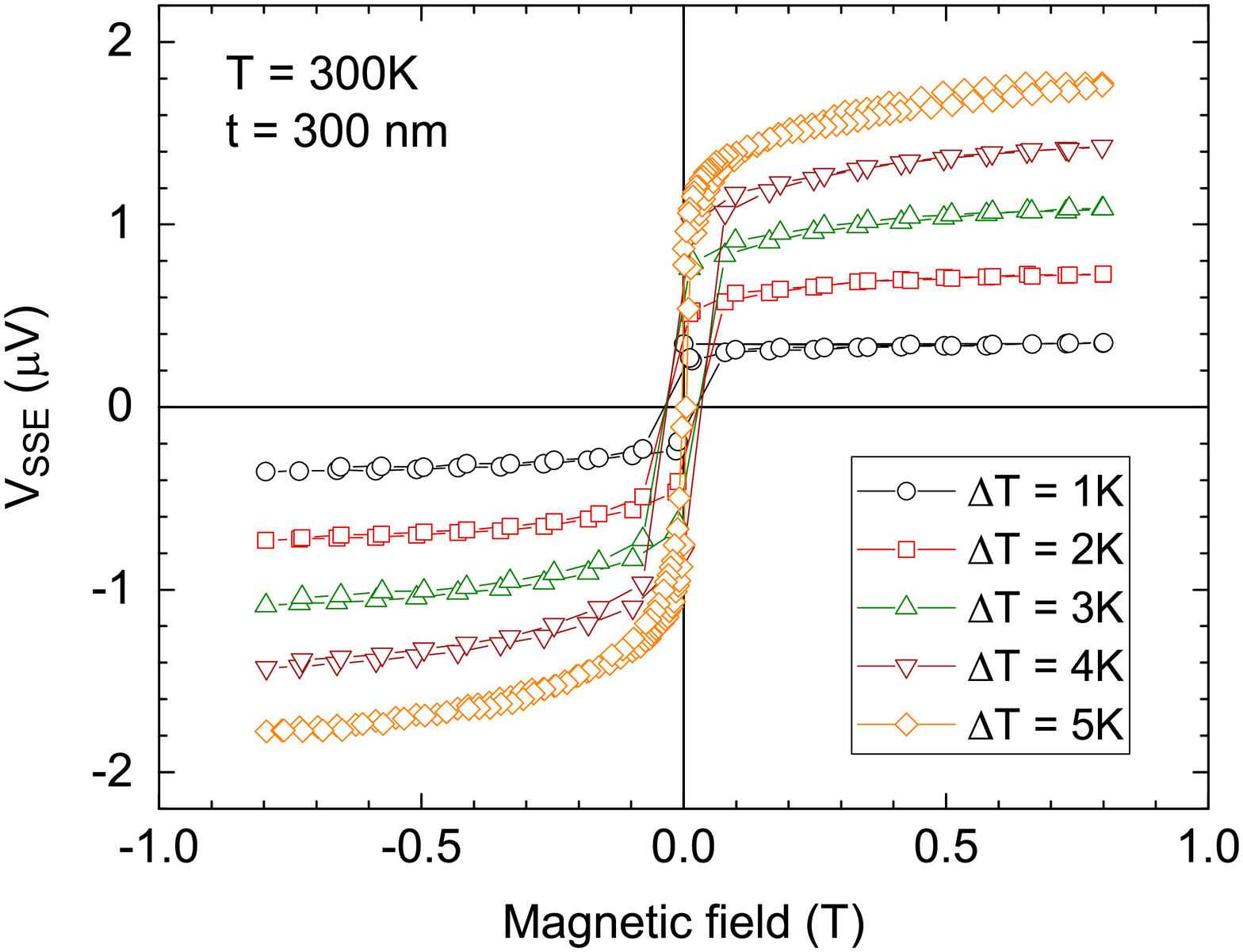} 
\includegraphics[width=0.85\columnwidth,viewport=0 0 772 591,clip]{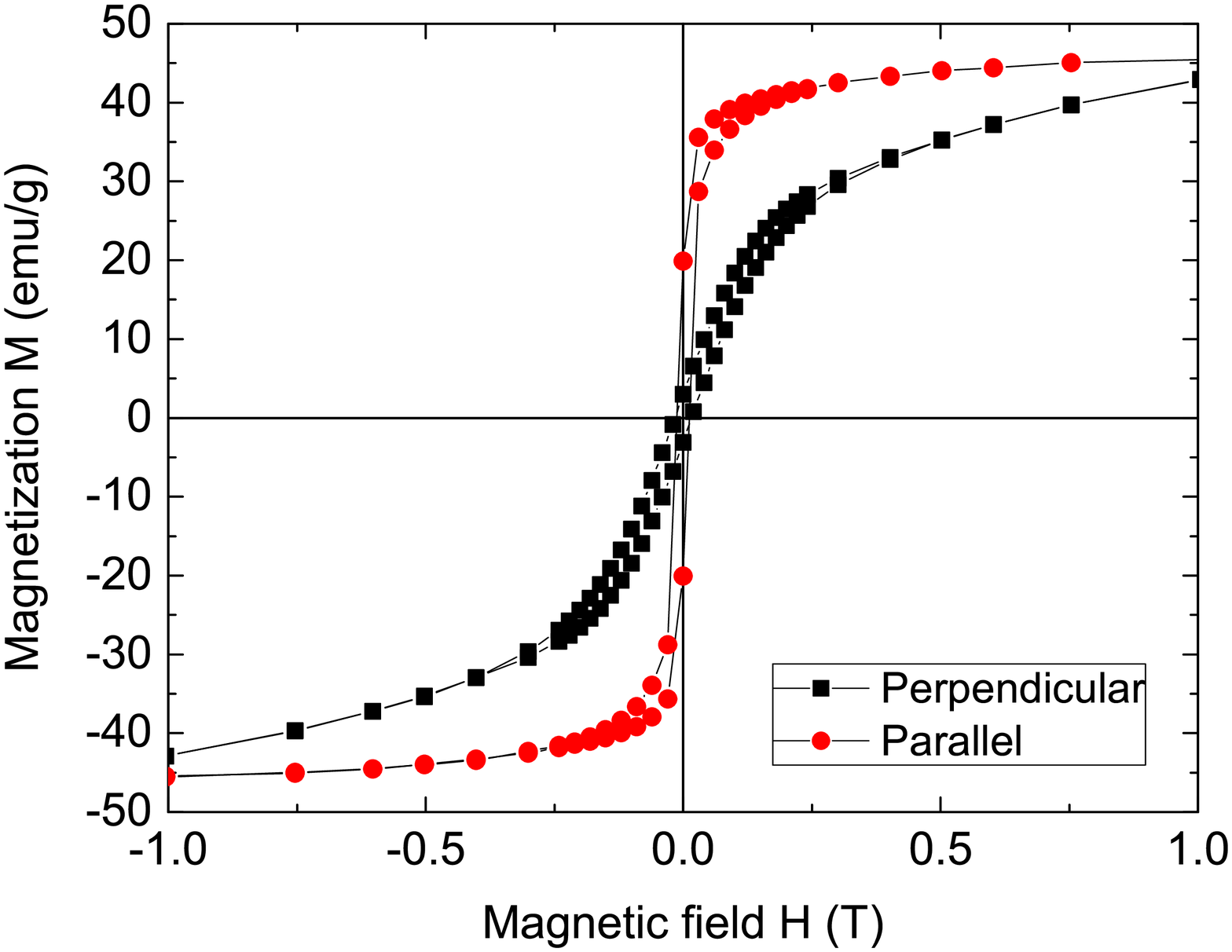} 
\caption{Spin Seebeck signal (upper panel), and in plane and out of plane magnetization (lower panel), in dependence on magnetic field at 300~K for \yhexzn.} \label{Fig_SSEmagZn}
\end{figure}


The magnetic properties of the Y-hexaferrites thin layers were characterized by magnetization curves measured at room temperature.
The magnetic moment of Zn2Y at $T_{room}$ determined from the saturated value of magnetization in parallel orientation is 11~\mub, see the lower panel of Fig.~\ref{Fig_SSEmagZn}, which is comparable with the expected value \cite{RefPullar2012PROGMS57_1191}. The measurement confirms that Zn2Y is a soft magnet with negligible hysteresis. The saturation in the orientation parallel with the thin layer is attained already at low field, whereas the saturation in the out of plane orientation, \textit{i.e.} along the $c$-direction, is achieved at higher field above $1\sim$~T, in agreement with the $ab$ easy plane orientation.


The Spin Seebeck signal of Zn2Y at room temperature is displayed in the upper panel of Fig.~\ref{Fig_SSEmagZn} for various temperature gradients applied across the thin layer. The measured voltage is positive in positive external magnetic field, in agreement with the positive spin Hall angle of Pt \cite{RefSinova2015REVMODPHYS87_1213}, and changes sign when switching the polarity of the magnetic field. The dependence on the magnetic field has the same shape with negligible hysteresis as the magnetization in parallel orientation. The data clearly show linear dependence on temperature gradient.

\begin{figure}
\centering
\includegraphics[width=0.85\columnwidth,viewport=0 0 772 591,clip]{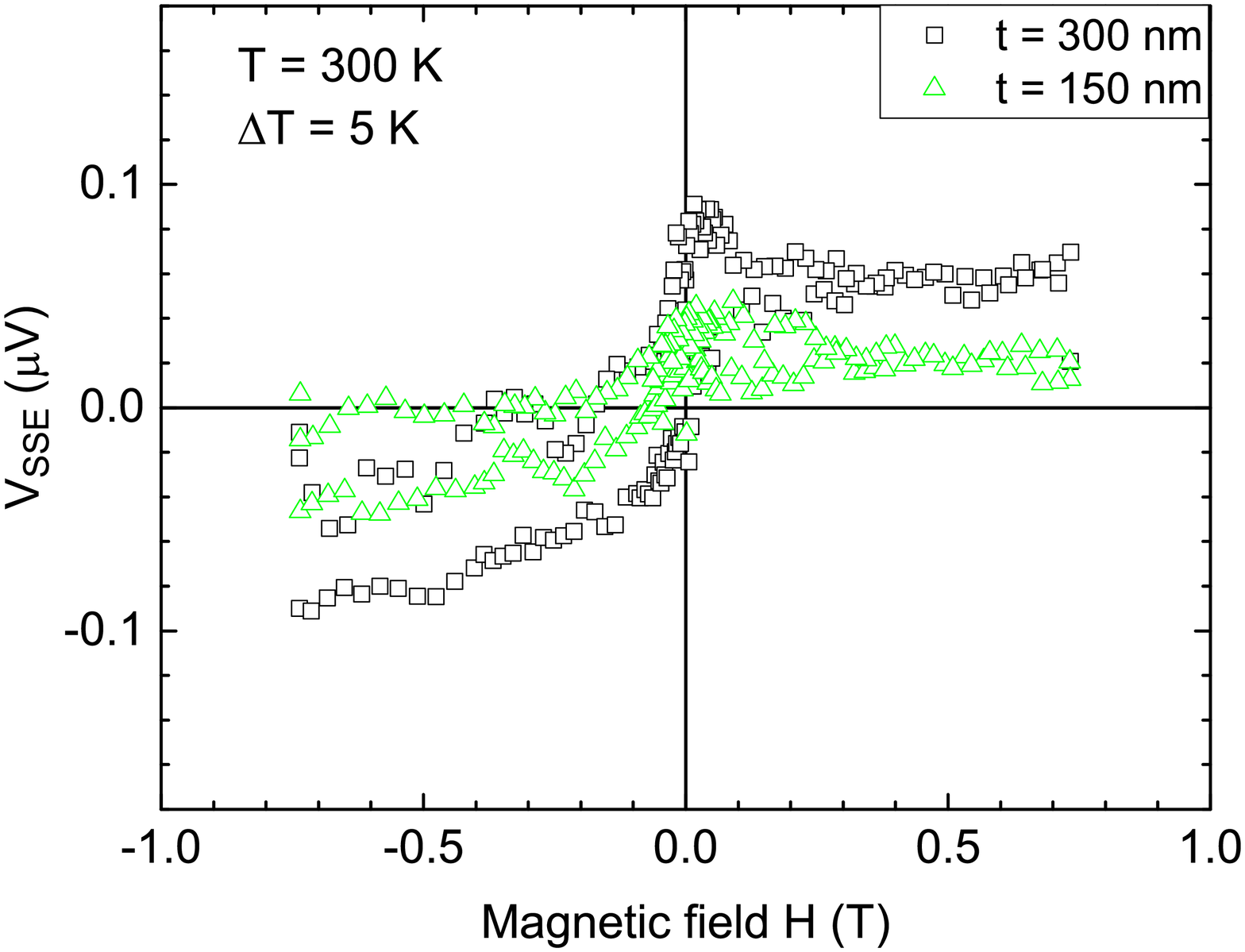} 
\includegraphics[width=0.85\columnwidth,viewport=0 0 772 591,clip]{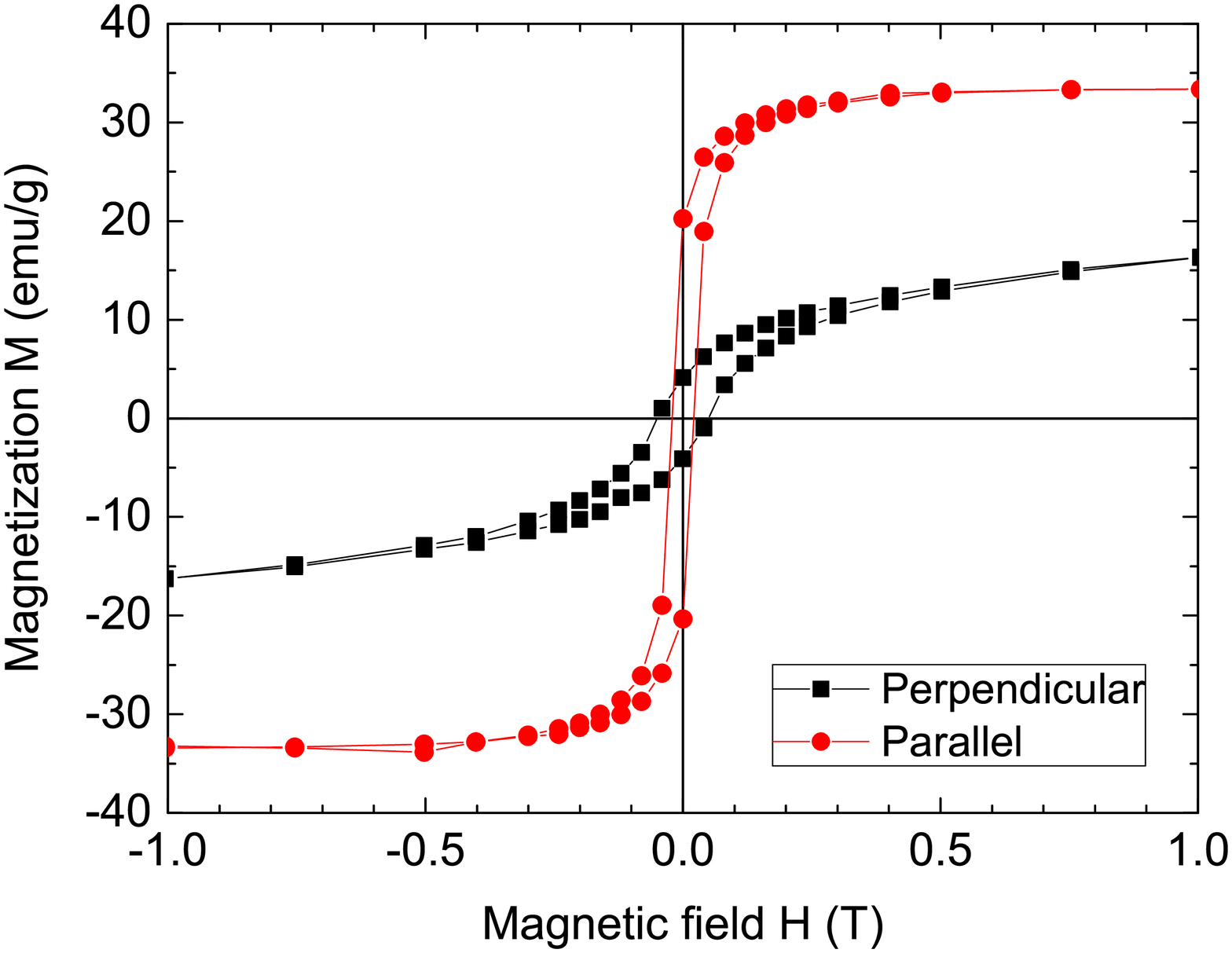} 
\caption{Spin Seebeck signal (upper panel) for 2 selected thin layers, and in plane and out of plane magnetization (lower panel), in dependence on magnetic field at 300~K for \yhexco.} \label{Fig_SSEmagCo}
\end{figure}


The magnetic moment of Co2Y at $T_{room}$ determined from the saturated value of magnetization in parallel orientation is 10~$\mu_B/f.u.$, see the lower panel of Fig.~\ref{Fig_SSEmagCo}. This value is slightly higher than the expected moment \cite{RefPullar2012PROGMS57_1191}, presumably due to relatively higher structural preference of Co for spin down sites in the case of our thin films. The difference between the saturation in parallel and perpendicular orientation is bigger in agreement with higher magnetocrystalline anisotropy of Co2Y compared to Zn2Y.


However, despite the similar magnetic properties, the SSE signal for Co2Y was not observed, see the upper panel of Fig.~\ref{Fig_SSEmagCo}.
To explain this different behavior of Zn2Y and Co2Y, we have considered the difference in the cation distributions of the transition metal cations over the structure, see for details the Introduction section. Since Zn$^{2+}$ ion is a non-magnetic, the structure of \yhexzn\ only contains one type of magnetic ion. Zn cation preferentially substitutes Fe$^{3+}$ in two tetrahedral sites with the same direction of spin polarization, therefore the total spin polarization of the unit cell does not significantly fluctuate across the material. In distinction, \yhexco\ contains two types of magnetic ions, \textit{i.e.} Co$^{2+}$ in low spin state in addition to Fe$^{3+}$ in high spin state. Co substitute Fe$^{3+}$ in all octahedral sites, where Fe may have both directions of spin polarization, without strong preference for a particular sites. We tentatively propose, that this random distribution of Co$^{2+}$ over octahedral sites with various spin polarizations interferes the long range magnetic ordering across the material, enhances the Gilbert damping constant and possibly results in suppressing of the SSE signal.


\begin{figure}
\centering
\includegraphics[width=0.99\columnwidth,viewport=0 0 830 580,clip]{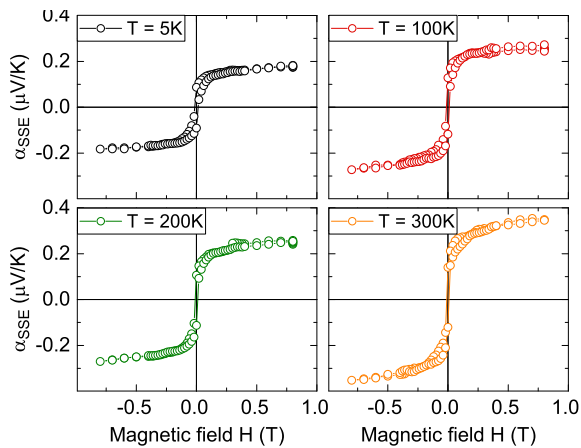} 
\caption{Spin Seebeck signal (SSE) dependence on magnetic field, divided by the total temperature difference $\triangle T$, for \yhexzn\ at selected temperatures.
 }\label{Fig_SSEvsTZn}
\end{figure}

SSE loops of \yhexzn\ was measured at several temperatures down to 5~K, see the measurements at selected temperatures 5, 100, 200 and 300~K in Fig.~\ref{Fig_SSEvsTZn}. The output power of the heater was the same for all temperatures.
The character of the loops is not changed with lowering temperature, the magnitude of the signal is only varying.

\begin{figure}
\centering
\includegraphics[width=0.85\columnwidth,viewport=0 400 560 770,clip]{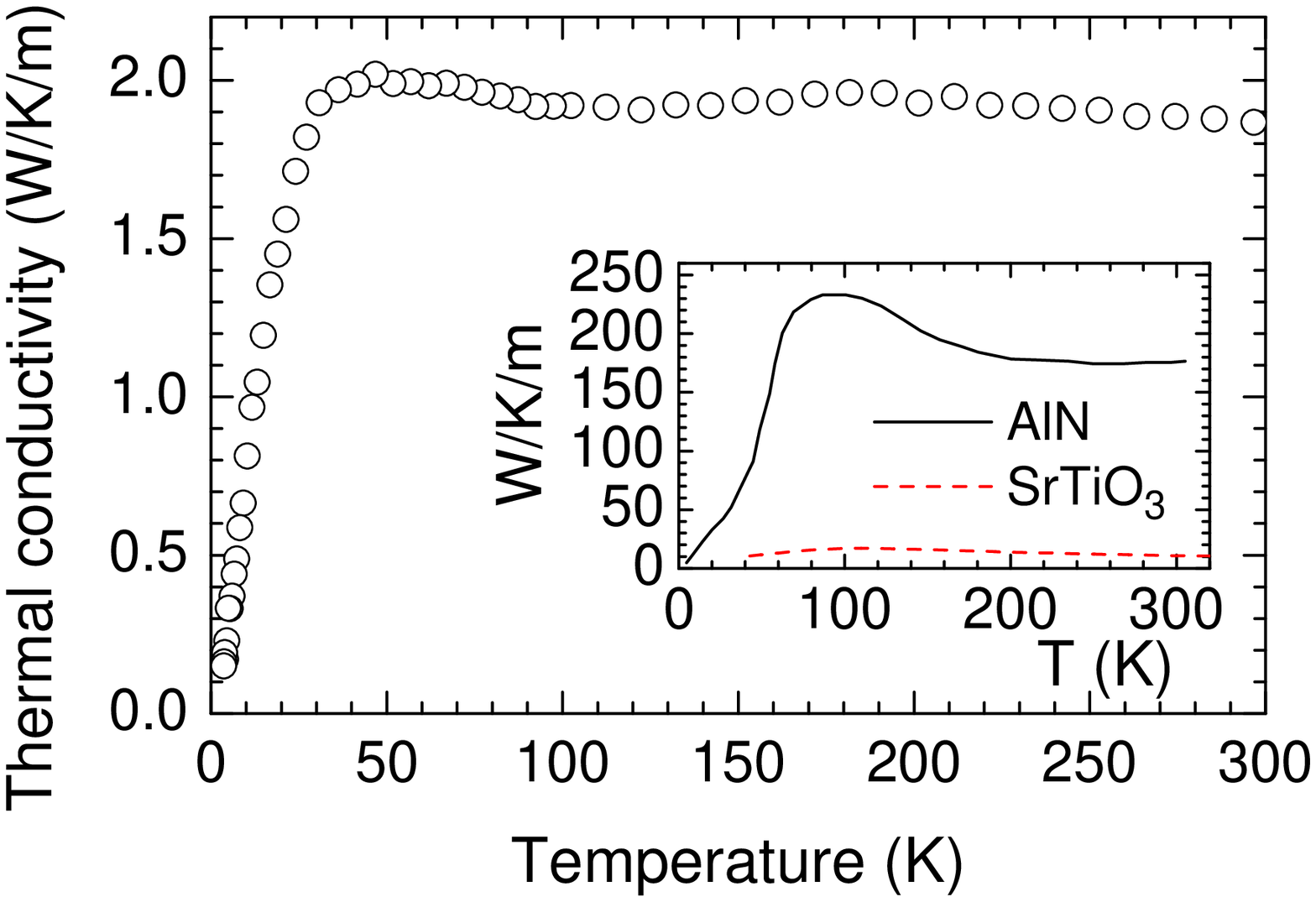} 
\caption{Thermal conductivity of bulk \yhexzn. Inset: thermal conductivity of AlN and SrTiO$_3$.} \label{Fig_ThcYhex}
\end{figure}


In order to investigate the temperature dependence of the SSE signal of Zn2Y in more details, we have performed measurement down to low temperature with 10~K step. The output power of the heater was also kept constant during this measurement. The value of the SSE was determined by switching the magnetic field to $\pm0.4$~T at each temperature and calculating the difference

\begin{equation}
SSE = \frac{V_{+0.4T} - V_{-0.4T}}{2}    \label{eqsseraw}
\end{equation}

\begin{figure}
\centering
\includegraphics[width=0.85\columnwidth,viewport=0 10 560 820,clip]{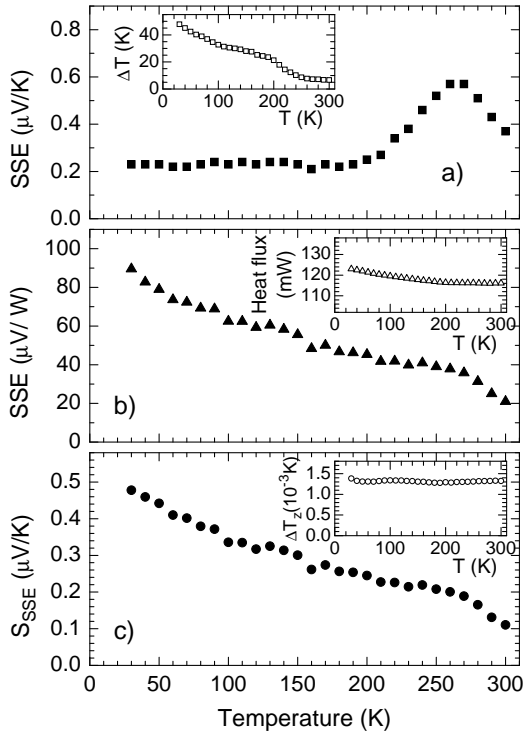} 
\caption{Spin Seebeck signal (SSE) dependence on temperature for \yhexzn, (a) divided by overall temperature gradient, inset: temperature difference. (b) divided by heat flux, inset: heat fluux. (c) calculated according to eq.~\ref{eqssenorm}, inset: temperature difference $\triangle T_{z}$.} \label{Fig_SSEthc}
\end{figure}

The resulting temperature dependence is displayed in the Fig.~\ref{Fig_SSEthc} in three ways. In the upper part of the figure, Fig.~\ref{Fig_SSEthc}a, the SSE signal is divided by the total temperature difference $\triangle T$ determined over the whole measuring cell. The temperature evolution of $\triangle T$ shown in the inset revealed, that $\triangle T$ increased several times during cooling. Since the output power of the heater was kept approximately constant, this increase should be related to a decrease of the thermal conductivity of the materials between the temperatures probes. To verify this assumption, we have measured thermal conductivity of the relevant materials, \textit{i.e.} the bulk sample \yhexzn\ synthesized from the precursors used for the thin layer deposition and compacted by isostatic pressing, AlN plate used to separate the heater and the sample, and the SrTiO$_3$ substrate, see Fig.~\ref{Fig_ThcYhex}. However, thermal conductivities of these materials weighted by their thickness in the measuring cell cannot explain the evolution of $\triangle T$. It is obvious, that in order to explain the observed temperature dependence of $\triangle T$, the thermal resistance of the thermal barriers between the attached parts of the cell must be taken into account. We have calculated, that the thermal resistance of the barriers at room temperature represents more than 50\% of the total thermal resistance of the cell, and its percentage increases with temperature.

Since the value of the total temperature difference $\triangle T$ cannot be used as independent variable in different measurement setups among various laboratories, another less setup dependent parameter should be used instead, in order to normalize the measured SSE signal. We used the heat flux through the sample, as it was proposed in Ref~\cite{RefSola2015JAP117_17C510}. SSE signal divided by the heat flux through the sample is displayed in Fig.~\ref{Fig_SSEvsTZn}b, the corresponding heat flux corrected for the heat losses due to radiation, is shown in the inset.

In order to extract quantity comparable over different measurement setups including the geometry of the sample, an expression for spin Seebeck effect related to sample dimensions and temperature difference over the sample itself was defined \cite{RefJaworski2010NATMAT9_898,RefSola2015JAP117_17C510}

\begin{equation}
S_{SSE} = \frac{V_{x} t_{z}}{d_{x} \triangle T_{z}}    \label{eqssenorm}
\end{equation}

where $V_{x}$ is the voltage measured, $t_{z}$ is the thickness of the magnetic material, $d_{x}$ is the electric contact distance, and $\triangle T_{z}$ is the temperature difference at the magnetic material along the thickness $t_{z}$, see Fig~\ref{Fig_ExpSSE}. With the knowledge of the heat flux and the thermal conductivity of the sample material \yhexzn\ we were able to calculate $S_{SSE}$ according to eq.~\ref{eqssenorm}, see the temperature dependence in Fig.~\ref{Fig_SSEthc}c, the evolution of $\triangle T_{z}$ is displayed in the inset.

The correct normalization of SSE signal is important not only for comparing among various measurement setups, but also for the correct determination of the temperature dependence, as it is evident by comparison of various temperature evolutions of SSE shown in Fig.~\ref{Fig_SSEvsTZn}. The SSE related to the total temperature difference $\triangle T$ (Fig.~\ref{Fig_SSEvsTZn}a) shows incorrect temperature dependence influenced by the temperature dependence of the total thermal conductivity of the measuring setup. We propose that the correct temperature dependence is determined by relating SSE to heat flux (Fig.~\ref{Fig_SSEvsTZn}b) or to temperature difference at the sample $\triangle T_{z}$ (Fig.~\ref{Fig_SSEvsTZn}c). In this case, SSE is almost linearly increasing with lowering temperature.

The almost 5$\times$ increase of SSE at low temperature compared to room temperature can be partially explained by the increased magnetization (almost 2$\times$), but the decrease of Gilbert damping factor $\alpha$ should be of greater influence in this regard. It was determined in the study of the temperature dependence of SSE signal in \yig\ garnet (YIG) \cite{RefGuo2016PRX6_31012}, that the effective propagation length of thermally excited magnons $\xi$ is proportional to $T^{-1}$, and since at the same time $\alpha \sim \xi^{-1}$ \cite{RefRitzmann2014PRB89_024409,RefRitzmann2015PRB92_174411}, it means that Gilbert damping factor $\alpha$, which is expected to suppress the SSE signal, is linearly decreasing with temperature.

In distinction to temperature dependence of SSE in YIG \cite{RefGuo2016PRX6_31012}, where a maximum in SSE was observed and explained by the interplay of the increase of magnon effective propagation length and decrease of the total number of thermally excited magnons, we observed no maximum down to low temperature. We ascribe it to the lower dispersion of acoustic branches in magnon spectra of Y-hexaferrite, which makes the influence of increasing total number of thermally excited magnons less important.

For the confrontation of the normalized room values between Y-hexaferrite and garnet, we have determined values 21~$\mu$V/W and $S_{SSE} = 0.11$~\muvk\ for \yhexzn, which are lower in comparison with 46.6~$\mu$V/W and 0.28~\muvk\ for \yig\ \cite{RefSola2015JAP117_17C510}, despite the higher magnetic moment of Zn2Y. We presume, that it is due to the lower Gilbert damping constant $\alpha$ and higher dispersion of acoustic branches in magnon spectra of \yig.

\section{Conclusions}

Spin Seebeck effect (SSE) has been investigated in thin films of two Y-hexagonal ferrites \yhexzn\ (Zn2Y) and \yhexco\ (Co2Y) deposited by spin-coating method on SrTiO$_3$(111) substrate. The SSE signal was observed for Zn2Y, whereas no significant SSE signal was detected for Co2Y. This can be explained by a presence of two different magnetic ions in Co2Y, whose random distribution over octahedral sites interferes the long range ordering and enhances the Gilbert damping constant. The magnitude of spin-Seebeck signal of Zn2Y normalized to the temperature difference at the investigated layer and sample dimensions ($S_{SSE}$) is comparable to the results measured on yttrium iron garnet \yig. $S_{SSE}$ of Zn2Y exhibits monotonically increasing behaviour with decreasing temperature, as a result of the simultaneous increase of the magnetization and magnon effective propagation length.

\textbf{Acknowledgement}.
This work was supported by Project No.~14-18392S of the Czech Science Foundation and SGS16/245/OHK4/3T/14 of CTU Prague.


\end{document}